\documentclass[aps,twocolumn,prb,showpacs,superscriptaddress]{revtex4-1}

\usepackage{graphicx}
\usepackage{dcolumn}
\usepackage{bm}
\usepackage{amsmath}
\usepackage{amssymb}
\usepackage{epsfig}
\usepackage{verbatim}
\usepackage{pstricks,pst-grad,pst-plot}
\usepackage{natbib}
\usepackage{url}

\begin{document}

\title{Capillary condensation in one-dimensional irregular confinement}

\author{Thomas~P.~Handford}
\affiliation{Department of Chemistry, University of Cambridge, Cambridge, UK}
\email{tph32@cam.ac.uk} 
\author{Francisco~J.~P{\'e}rez-Reche}
\affiliation{Institute for Complex Systems and Mathematical Biology, SUPA, University of Aberdeen, Aberdeen, UK }
\email{fperez-reche@abdn.ac.uk}
\author{Sergei~N.~Taraskin}
\affiliation{St. Catharine's College and Department of Chemistry,
University of Cambridge, Cambridge, UK}
\email{snt1000@cam.ac.uk}

\begin{abstract}
A lattice-gas model with heterogeneity is developed for the description of fluid condensation in finite sized one-dimensional pores of arbitrary shape.
An exact solution of the model is presented for zero-temperature that reproduces the experimentally observed dependence of the amount of fluid adsorbed in the pore on external pressure. 
Finite-temperature Metropolis dynamics simulations support analytical findings in the limit of low temperatures.
The proposed framework is viewed as a fundamental building block of the theory of capillary condensation necessary for reliable structural analysis of complex porous media from adsorption-desorption data.
\end{abstract}

\pacs{75.10.Nr, 05.70.Np, 68.43.-h, 64.70.F-}

\maketitle

\section{Introduction}

Physical systems which consist of networks of pores, such as Vycor~\cite{Woo2003}, Silica aerogels~\cite{Gelb1999}, porous rocks~\cite{Broseta2001}, soil~\cite{PerezRechePRL2012} and others, 
have a wide spectrum of applications, ranging from
molecular filters and catalysts~\cite{Ravikovitch1997} to fuel storage~\cite{Felderhoff2007}.
Capillary condensation is an important and peculiar physical phenomenon occurring in many such systems.
Depending upon the pore structure and the materials involved, the adsorbed fluid density can exhibit both hysteresis and avalanches (abrupt changes in density) as the pressure varies.
In recent years, a lot of experimental and theoretical work has been undertaken in order to understand this dependence~\cite{Gelb1999,Valiullin2009,Monson2012}.
However, 
the understanding is not fully complete~\cite{Monson2012} and, for example,
routinely used classical theories fail to explain the appearance and shape of hysteresis in sorption curves of capillary condensation even in simple one-dimensional (1D) isolated pores~\cite{Evans1990,Coasne2001,Wallacher2004}.
Indeed, according to classical theories, pores closed at one end should not have hysteresis.
In contrast, the adsorption of $N_2$ into MCM-41 mesoporous silica consisting of pores closed at one end at the temperature $T=78$K reveals a hysteresis loop of the so-called H2-type~\cite{Sing1985}, i.e. involving a smooth increase in density for adsorption and sharp drop in density for desorption. 
An accurate theory for capillary condensation in 1D pores is ultimately necessary to ensure that, for instance, this phenomenon can be used as a reliable technique to probe the structure of generic porous media consisting of a network of interlinked 1D pores~\cite{Yortsos_1999,Gelb1999}.
 
A number of techniques have been developed to study capillary condensation theoretically, 
including microscopic molecular dynamics~\cite{ValiullinNature2006}, density functional theory~\cite{Evans_JChemPhys1986,Malijevsky_JChemPhys2012} and lattice-gas mean-field theory~\cite{Kierlik1998,Kierlik2001,Woo2003}.
Such studies have been conducted on a variety of different porous media. 
For one-dimensional pores, 
mean-field theory~\cite{Naumov2008} and multi-scale molecular dynamics studies~\cite{Puibasset2007,Puibasset2009} have previously been used to test 
the hypothesis that hysteresis is caused by 
heterogeneity~\cite{Wallacher2004}, which includes variations in pore diameter, chemical heterogeneity in the pore walls and roughness of internal surfaces~\cite{Maddox1997,Edler1998,Fenelonov2001,Sonwane2005}.
Additionally, these numerical studies have revealed the occurrence of avalanches in the amount of adsorbed fluid during condensation and evaporation.
Such avalanches bare similarity to avalanches in magnetisation found in the random-field Ising model (RFIM)~\cite{Sethna1993,Kierlik1998,Woo2003,PerezRechePRL2005}.
Employing this similarity, we map RFIM to the lattice-gas model and demonstrate that: 
(i) a heterogeneous lattice-gas model is a minimally sufficient model to reproduce experimental observations of variations of fluid density with pressure in finite-sized pores; 
(ii) this model can be solved exactly analytically at zero temperature ($T=0$) by a novel technique, with the solution being fully supported by numerical simulations; 
(iii) such an analytical solution leads to simple physical explanations and interpretations of experimental results for condensation in 1D pores that remain qualitatively valid at sufficiently low finite temperatures. 

More specifically, our findings are as follows.
(i) The effects of the closed and open ends of the pores, considered important in the classical theories, are significantly reduced for large disorder in strengths of the interactions of the fluid with the pore walls.
(ii) A positive-skewed distribution in strengths of such interactions can lead to sorption curves of the same form as those found experimentally~\cite{Coasne2001,Wallacher2004}.
(iii) It is predicted that the mechanism for adsorption depends crucially on the length of pores and their geometry. For short pores, adsorption isotherms may depend on whether pores are open or close at the ends. 
In contrast, the dependence of adsorption on the characteristics at the pore ends is lost for long enough pores.
(iv) Sorption is shown to have several different mechanisms, leading to different forms of sorption curves. 
However, sorption in long pores, or pores characterised by a large degree of heterogeneity, shows universal features typical of a disorder-controlled regime. 
(v) In cylindrical pores consisting of two parts of different diameter, condensation and evaporation in one part can induce condensation and evaporation in the rest of the pore for low disorder, but for high disorder, the two parts of the pore behave independently. 

The structure of this paper is as follows. 
The model for condensation and pore geometries studied are introduced in Sec.~\ref{Sec:Model}. 
Sec.~\ref{Sec:Exact_Solution} presents the exact solution of the proposed model at zero temperature. 
The results for adsorption/desorption are presented in sections \ref{Sec:Results_T0} and \ref{Sec:Results_Tgt0} for $T=0$ and $T>0$, respectively. 
Finally, the conclusions are given in Sec.~\ref{Sec:Conclusions}.

\section{Model and pore geometries}
\label{Sec:Model}

\subsection{Lattice-gas model with disorder in matrix-fluid interactions}
The proposed model is based on a standard lattice-gas model of capillary condensation~\cite{Kierlik1998,Gelb1999,Evans1990,Woo2003}.
In this model, the 3D space is split into cells, and each cell can be filled either by matrix (the solid substrate), liquid or vapour. 
If the cell $i$ is occupied by matrix (expressed by setting parameter $\eta_i=0$), then it cannot become occupied by fluid. 
It is assumed that the variables $\eta_i$ are quenched for the whole system and thus the matrix state cannot change during the condensation. 
If cell $i$ is not occupied by the matrix ($\eta_i=1$), then it can be occupied by either liquid ($\tau_i=1$) or vapour ($\tau_i=0$).
The variables $\tau_i$ can vary with the change in chemical potential, $\mu$.
The Hamiltonian which describes the lattice gas model is given by~\cite{Kierlik1998,Kierlik2001,Woo2003},
\begin{eqnarray}
{\cal H}&=&-\mu\sum_i\tau_i\eta_i-w^{\text{ff}}\sum_{\langle ij\rangle}\tau_i\tau_j\eta_i\eta_j\nonumber\\
&-&\sum_{\langle ij\rangle}\left[\tau_iw^{\text{mf}}_{ij}\eta_i(1-\eta_j)+\tau_jw^{\text{mf}}_{ji}\eta_j(1-\eta_i)\right]~,\label{eq:Hamiltonian}
\end{eqnarray}
where the summations run over all the cells in the system in the first term and over all nearest-neighbour pairs $\langle ij\rangle$ in the other terms. 
The fluid-fluid interaction parameter, $w^{\text{ff}}>0$, is 
assumed to be the same for all pairs of cells.
The matrix-fluid interaction strength between the matrix at cell $j$ and fluid at the neighbouring cell $i$ is described by the parameter $w^{\text{mf}}_{ij}$.
The values of $w^{\text{mf}}_{ij}$ are considered to be independently distributed quenched random variables with the probability density function, $\rho_i(w^{\text{mf}}_{ij})$, which can be cell dependent.
The random distribution of this parameter has been studied previously in the context of chemical heterogeneity of the pore walls~\cite{Naumov2009}, but below it is assumed to characterise all types of heterogeneity. 
For concreteness, we focus on two forms of disorder in $w^{\text{mf}}_{ij}$, representing heterogeneity on different scales ranging from local fluctuations at a single point on the pore wall to a variable diameter of the pore. 
More specifically, we consider (i) a normal distribution, 
\begin{equation}
\rho_i(w^{\text{mf}}_{ij})={\cal N}(\langle w^{\text{mf}}\rangle_i,\Delta_i^2)~,\label{eq:gaussDistribution}
\end{equation}
and (ii) an exponential distribution with correlations ensuring that all $w^{\text{mf}}_{ij}$ are the same for the same cell $i$ (i.e. $w^{\text{mf}}_{ij}$ is independent of $j$), and distributed according to, 
\begin{equation}
\rho_i(w^{\text{mf}}_{ij})=\Theta(x)\Delta_i^{-1}\exp(-\Delta_i^{-1}x)~,\label{eq:expDistribution}
\end{equation}
where $x=(w^{\text{mf}}_{ij}-\langle w^{\text{mf}}\rangle_i)+\Delta_i$ and $\Theta(x)$ is the Heaviside step function.
In Eqs.~\eqref{eq:gaussDistribution} and~\eqref{eq:expDistribution}, $\langle w^{\text{mf}}\rangle_i$ and $\Delta_i^2$ refer to the mean and variance, respectively, of the corresponding distributions.
Gaussian uncorrelated heterogeneity describes local fluctuations of the matrix-fluid interaction. In contrast, correlations in the exponential heterogeneity are introduced to describe heterogeneity in pore diameter.

\subsection{Mapping to the random-field Ising model}\label{sec:Mapping}

The lattice gas model can be mapped to the RFIM~\cite{Kierlik1998,Woo2003} which gives an intuitively simpler representation and is technically more convenient for the calculations presented below. In the RFIM representation, the Hamiltonian described by Eq.~\eqref{eq:Hamiltonian} is given by,
\begin{equation}
{\cal H}=-J\sum_{\langle ij \rangle}s_is_j-\sum_ih_is_i-H\sum_is_i~,
\end{equation}
where the variable $s_i=(2\tau_i-1)\eta_i$ ($s_i=\pm 1$) represents a spin state, $H=\mu/2$ refers to an external magnetic field,
and $J=w^{\text{ff}}/4$ describes the spin-spin interaction.
The fields $h_i$ at cell $i$ are given by 
\begin{equation}
h_i=\sum_{j/i}\left[(1-\eta_j)\frac{w^{\text{mf}}_{ij}}{2}+\eta_j\frac{w^{\text{ff}}}{4}\right]~,\label{eq:RandomField}
\end{equation}
with the sum running over all nearest neighbours of $i$. 
Since $w^{\text{mf}}_{ij}$ are randomly distributed, the fields $h_i$ defined by~\eqref{eq:RandomField} are also random variables, with distribution $\rho_{h}(h_i)$ which depends on the pore geometry chosen. 
More explicitly, $\rho_{h}(h_i)$
depends on the numbers, $n^{\text{m}}_i=\sum_{j/i}(1-\eta_j)$ and $n^{\text{u}}_i=\sum_{j/i} \eta_j$, of neighbouring cells which are occupied and unoccupied by matrix, respectively. 
When $w^{\text{mf}}_{ij}$ is normally distributed, the random field at cell $i$ is distributed according to $\rho_{h_i}(h_i)={\cal N}(\langle h_{i}\rangle_i,{\Delta_{h_i}}^2)$ with $\langle h_{i}\rangle_i=(n^{\text{m}}_i/2)\langle w^{\text{mf}}\rangle_i+(n^{\text{u}}_i/4)w^{\text{ff}}$ and $\Delta_{h_i}^2=(n^{\text{m}}_i/2)\Delta_i^2$.
When $w^{\text{mf}}_{ij}$ follows a correlated exponential distribution, the random field at cell $i$ is distributed according to 
$\rho_{h_i}(h_i)=\Theta(y)\Delta_{h_i}^{-1}\exp(-\Delta_{h_i}^{-1} y)$ (where $y=h_i-\langle h_i \rangle_i+\Delta_{h_i}$), with the mean $\langle h_{i}\rangle_i$ being the same as for the normal distribution and standard deviation $\Delta_{h_i} = n_i^\text{m} \Delta_i/2$.
In both cases, the values of $\langle h_{i}\rangle_i$ and ${\Delta_{h_i}}^2$ can differ between the cells.

\subsection{Pore geometries}

\begin{figure}
\includegraphics[width=7.0cm]{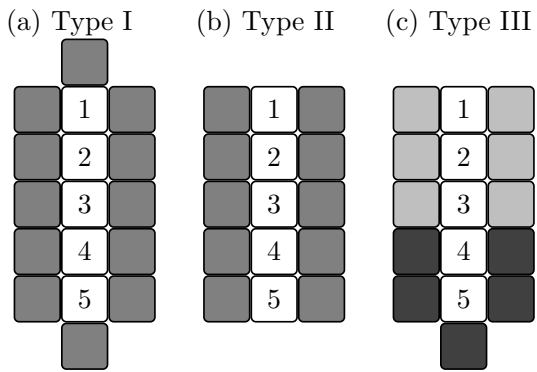}
\caption{Diagrams showing 2D sections of linear pores of $N=5$ cells unoccupied by matrix, labelled $i=1,\ldots,5$ (white squares), and different configurations of matrix cells (grey squares) in the 3D system. (a) Pore of type I completely surrounded by matrix. (b) Pore of type II with open ends. (c) Pore of type III with one open and one closed end consisting of two sections with statistically different matrix-fluid interaction represented by light- and dark-grey cells. 
The interaction of the fluid in cells $1\le i\le N_1=3$ and in cells $4< i\le N$ with the matrix represented by light-grey and dark-grey squares is described by $\langle w^{\text{mf}}_1\rangle$, and $\langle w^{\text{mf}}_2\rangle$, respectively.
For clarity, 2 relevant matrix cells per unoccupied cell, on top of and underneath the white cells, are not shown in this 2D diagram. \label{fig:Cell}}
\end{figure}

Three types of
1D pore geometries consisting of $N$ cells embedded in a simple cubic lattice are analysed below.
They are linear pores of (i) type I with both ends closed by matrix (see Fig.~\ref{fig:Cell}(a)), (ii) type II with both ends open and bounded by vapour (see Fig.~\ref{fig:Cell}(b)), and (iii) type III with one open end and one closed end. 
In the first two cases, we consider $w^{\text{mf}}$ to be identically and independently distributed random variables with $\langle w^{\text{mf}}\rangle_i=\langle w^{\text{mf}}\rangle$ and $\Delta_i=\Delta$.
In the last case, we consider two distinct sections of lengths $N_1$ and $N-N_1$ characterised by different values of mean matrix-fluid interaction $\langle w^{\text{mf}}_1\rangle$ and $\langle w^{\text{mf}}_2\rangle$, but the same variance $\Delta$, respectively (see Fig.~\ref{fig:Cell}(c)). 
Pores of type III are intended to model ink-bottle and funnel pore geometries, with weaker matrix-fluid interaction representing a larger diameter.

\subsection{Dynamics}\label{sec:Dynamics}

We study adsorption and desorption isotherms obtained by sweeping the chemical potential at a rate $r$ from $\mu=-\infty$ to $\infty$ and back again. 
The state of the system is described in terms of the mean volume of the absorbed liquid, $\langle V \rangle=\left\langle N^{-1}\sum_{i}\tau_i\right\rangle$ and the variance of this quantity, 
$\text{Var}[V]=\left\langle \left(N^{-1}\sum_{i}\tau_i\right)^2\right\rangle-\langle V\rangle^2$. 
When the system is driven in this way, it evolves through a rugged free energy landscape corresponding to the Hamiltonian ${\cal H}$ which, due to the presence of random fields, consists of an exponentially large number of local minima for given $\mu$ and $\langle V \rangle$~\cite{Woo2003,Puibasset_JPhysC2011,PerezReche_PRB2008}. 
The evolution of the system (i.e. changes of the occupation numbers, $\{\tau_i\}$) 
is caused either by changes in the energy landscape due to variations of $\mu$, or by thermal effects.
Below the critical temperature, $T_c$, where condensation occurs in a discontinuous manner~\cite{Gelb1999}, random fields introduce glass-like behaviour to the system and thermally activated transitions are unlikely on the time-scale of real experiments~\cite{Woo2003}. 
That is why the results of mean-field lattice-gas models~\cite{Kierlik2001,Woo2003}, which ignore thermal fluctuations, and experiments~\cite{Gelb1999} are in good agreement.
Temperature is known to affect hysteresis~\cite{Gelb1999} but this has been properly accounted for by mean-field theories in terms of entropic contributions that represent a quenched modification to the energy landscape~\cite{Kierlik2001,Woo2003}.

Motivated by these results, we first obtain the exact analytical solution to the proposed model at zero temperature and then study the effects of non-zero temperature by means of Monte-Carlo numerical simulations. 
These simulations include thermally activated events but their effect is expected to play a secondary role on hysteresis at low temperatures and are not analysed in detail.

At $T=0$, we employ a single-spin flip metastable dynamics that has been widely used within the context of the zero-temperature RFIM (zt-RFIM)~\cite{Sethna1993}. 
According to this dynamics, for adsorption (desorption) each cell is initially empty (occupied), $s_i=-1$ ($s_i=+1$), and can become occupied by liquid (gas), $s_i=+1$ ($s_i=-1$), once its local field,
\begin{eqnarray}
f_i&=&H+h_i+J\sum_{j/i}s_j\nonumber\\
&=&\frac{\mu}{2}+\sum_{j/i}(1-\eta_j)\frac{w_{ij}^{\text{mf}}}{2}+\frac{w^{\text{ff}}}{2}n_i^f~, \label{eq:LocalField}
\end{eqnarray}
is positive, $f_i>0$ (negative, $f_i<0$). 
Here, $n_i^\text{f} = \sum_{j/i}\eta_j\tau_j$ is the number of fluid cells neighbouring cell $i$. 
According to the above rule, a configuration of phases is stable if all the fluid cells satisfy the condition $s_if_i >0$. 
The system is driven quasistatically under the assumption that the rate of relaxation of spins is much larger than the rate $r$ of variation of $\mu$~\cite{Sethna1993,PerezRechePRL2005}. 
In practise, this is achieved by 
sweeping $\mu$ until at least one spin becomes unstable. 
At this point, an avalanche starts and $\mu$ is kept constant until a new stable configuration is reached. 
After that, activity can only resume if $\mu$ is varied and a new avalanche is induced by this variation.

Adsorption/desorption processes at $T>0$ are simulated numerically using Metropolis dynamics implemented as follows~\cite{LandauBOOK}. 
For given $\mu$,
a cell is chosen at random and a change of its state, $\tau_i$ (or $s_i$), is proposed. 
The change is accepted with probability $p=\min \{1,\exp(-\beta\Delta {\cal H})\}$, where $\beta = (k_B T)^{-1}$ and $\Delta {\cal H}=f_is_i$ is the change in energy of the proposed change of state.
This process is referred to as a single Monte Carlo step, 
with the unit of time in the simulation being Monte-Carlo Steps per Spin (MCSS). 
As the simulation progresses the value of $\mu$ is incremented at a fixed rate $r$ measured in MCSS$^{-1}$ so that the increment at each step is $\delta\mu=r/N$.
This dynamics reduces to the $T=0$ dynamics in the limit of $r\to 0$ and $\beta\to \infty$.

\section{Exact solution of the model at $T=0$}
\label{Sec:Exact_Solution}
In this section, we obtain exact analytical solutions for
sorption curves for a 1D pore. 
In order to do that, we develop a novel method based on the use of the generating function formalism~\cite{Fisher1961,Sabhapandit2000}.
Within this formalism, the volume of fluid in the finite pore is a random value and can be expressed by a generating function,
\begin{equation}
G(x)=\sum_{n=0}^NP(n)x^n~, \label{eq:GenFunc}
\end{equation}
where $P(n)$ is the probability that $n=\sum_{i=1}^N\tau_i$ cells in the finite pore are occupied by fluid.
The mean volume of fluid is given by,
\begin{eqnarray}
\langle V\rangle=V_0\langle n\rangle=\partial_xG(1)~,\label{eq:MeanFromGen}
\end{eqnarray}
with volume of a cell set to $V_0=1$, 
and the variance is,
\begin{eqnarray}
\text{Var}[V]&=&\partial_{xx}G(1)+\partial_xG(1)-\left[\partial_{x}G(1)\right]^2~,\label{eq:VarFromGen}
\end{eqnarray}
where $\partial_{x}G(1)$ and $\partial_{xx}G(1)$ refer to the first and second derivatives of $G(x)$ with respect to $x$ evaluated at $x=1$.
We derive a form for these expression by establishing a recursion relation for $G(x)$ in the following way.

We assume particular sequences for occupancy of cells that are convenient from a mathematical viewpoint. This is possible owing to the
abelian property of the zt-RFIM which ensures that the final state of the system is independent of the order in which the cells are occupied by fluid (the order spins flip)
as long as the system ends up in a stable or metastable state~\cite{Sethna1993,Dhar1997}.
The sequence we chose assumes that the relaxation of the system into a metastable state takes place in a series of $N$ time-steps, $t$, $1\le t\le N$.
Initially, during time-step $t=1$, all the cells with $i\ge 2$ are artificially prevented from being occupied, while the first cell $i=1$ is allowed to change its state.
Cell $i=1$ can either change state from unoccupied to occupied if the local field given by Eq.~\eqref{eq:LocalField} is positive, $f_1>0$, or remain unoccupied if the local field is negative. 
This process with two possible outcomes is called relaxation of cell $1$.
Next, during time-step $t=2$, we allow cell $i=2$ to relax, while cells $i\ge 3$ are still held in the unoccupied state, and cell $i=2$ can become occupied if $f_2>0$.
If cell $i=2$ does become occupied, then the local field $f_1$ at cell $i=1$ will increase.
This can cause cell $i=1$ to become occupied if it was not occupied already, i.e. an avalanche can pass from cell $i=2$ to cell $i=1$ during time-step $t=2$.
Similarly, we allow the next cell in the pore, i.e. $i=3$, to relax and if it becomes occupied an avalanche can pass back along the pore towards cell $i=1$ if those cells with $i<3$ were not occupied.
This method is recursively applied $N$ times until all the cells in the system are relaxed.

Let us consider the cell $i=N$ at the end of the pore, which is the last cell to be allowed to relax in the above procedure.
At the start of time-step $t=N$, the neighbouring pore cell $i=N-1$ can be occupied or unoccupied, i.e. $s_{N-1}=\pm 1$. 
If the neighbouring cell $i=N-1$ is occupied ($s_{N-1}=+1$), and the random field at cell $i=N$ is $h_N>-J-H$, then the local field $f_N>0$ and cell $i=N$ will become occupied.
This occurs with cell-dependent probability $p^{\prime}_{N,1}$ where, 
\begin{equation}
p_{i,m}^{\prime}=\int_{h=-H-J(2m-1)}^\infty \rho_{h_i}(h)\text{d}h~,\label{eq:pmEnd}
\end{equation}
with $m$ being the number of occupied neighbours of cell $i=N$, i.e. $m=1$ in this case.
If, however, the neighbouring cell $i=N-1$ is unoccupied ($s_{N-1}=-1$), then the local random field must be above a higher threshold for cell $N$ to become occupied, $h_N>J-H$, which occurs with probability, $p^{\prime}_{N,0}$, given by Eq.~\eqref{eq:pmEnd} with $m=0$. 
In this case, the field at cell $i=N-1$ will increase, and an avalanche can propagate back along the pore.

The probability that there are $n$ occupied cells in the lattice at the end of the relaxation process can therefore be written as,
\begin{eqnarray}
&&P(n)=P_{A}(N-1,n-1)p^\prime_{N,1}\nonumber\\
&&+P_{A}(N-1,n)(1-p^\prime_{N,1})+P_{B}(N-1,n-1)p^\prime_{N,0}\nonumber\\
&&+ P_{C}(N-1,n)(1-p^\prime_{N,0})~,\label{eq:ProbEnd}
\end{eqnarray}
in terms of the probabilities, $P_{A}(i,n^\prime)$, $P_{B}(i,n^\prime)$ and $P_{C}(i,n^\prime)$, which can be recursively determined.
The quantity, 
\begin{equation}
P_{A}(i,n^\prime)=\text{Prob}[n_i(t=i)=n^\prime \cap s_{i}(t=i)=+1]~,\label{eq:ProbA}
\end{equation}
is the probability that at the end of time-step $t=i$, there are $n_i(t=i)=n^\prime$ occupied cells with index $j$ in the range $1\le j \le i$ (i.e. $n_i=\sum_{j=1}^{i}(s_j+1)/2$)
 and that cell $i$ is occupied, $s_{i}(t=i)=+1$.
The value, 
\begin{eqnarray}
P_{B}(i,n^\prime)=&&\text{Prob}[s_{i}(t=i)=-1 \nonumber\\&&\cap n_i(t=N)=n^\prime|s_{i+1}(t=N)=+1]~,\label{eq:ProbB}
\end{eqnarray}
is the probability that cell $i$ is unoccupied at end of time-step $t=i$, $s_{i}(t=i)=-1$, and at the end of the relaxation process (end of time-step $t=N$) there are $n_i=n^\prime$ occupied cells in the range $1\le j \le i$, given that cell $i+1$ becomes occupied, $s_{i+1}(t=N)=+1$, during some time-step $t^\prime$, $i<t^\prime\le N$ (which causes avalanches to pass back along the pore towards cell $1$, changing the occupation number $n_i$).
The third quantity,
\begin{equation}
P_{C}(i,n^\prime)=\text{Prob}[n_i(t=i)=n^\prime \cap s_{i}(t=i)=-1]~,\label{eq:ProbC}
\end{equation}
is the probability that at the end of time-step $t=i$ there are $n_i(t=i)=n^\prime$ occupied cells in the range $1\le j \le i$ and that cell $i$ is unoccupied at this time-step, $s_{i}(t=i)=-1$.

Using Eqs.~\eqref{eq:GenFunc} and~\eqref{eq:ProbEnd}, the generating function $G(x)$ for the total number of occupied cells in the pore can be written in terms of the generating functions $A_{i}(x)$, $B_{i}(x)$ and $C_{i}(x)$ for the corresponding probabilities defined by Eqs.~\eqref{eq:ProbA}-\eqref{eq:ProbC} as,
\begin{eqnarray}
G(x)&=&\left[xp^\prime_{N,1}+(1-p_{N,1}^\prime)\right]A_{N-1}(x)\nonumber\\&+&xp_{N,0}^\prime B_{N-1}(x)+(1-p^\prime_{N,0})C_{N-1}(x)~.\label{eq:GenEnd}
\end{eqnarray}
where the generating function $A_{i}(x)$ is defined as
$A_{i}(x)=\sum_{n=0}^{i}P_{A}(i,n)x^n$,
while $B_{i}(x)$ and $C_{i}(x)$ are defined according to the same relation with $A$ replaced by $B$ and $C$, respectively.

Expressions for the generating functions $A_{i}(x)$, $B_{i}(x)$ and $C_{i}(x)$ for $i>1$ can be found recursively in the following way.
If cell $i-1$ is occupied at the start of time-step $t=i$ and the local field at cell $i$ is positive, i.e. $f_i=h_i+H>0$, then cell $i$ will become occupied during time-step $t=i$. 
This occurs with probability $p_{i,1}$, where 
\begin{equation}
p_{i,m}=\int_{h=-H-J(2m-2)}^\infty \rho_{h_i}(h)\text{d}h~.\label{eq:pmMiddle}
\end{equation}
If cell $i-1$ is unoccupied then the random field has to be at a higher threshold, $h_i>-H+2J$, in order for cell $i$ to become occupied during time-step $i$.
The random field will be above this higher threshold with probability $p_{i,0}$, given by Eq.~\eqref{eq:pmMiddle}. 
If cell $i$ does become occupied during time-step $t=i$ and $s_{i-1}=-1$ at this time, then an avalanche might pass back along the pore towards cell $i=1$.
This gives the expressions for $P_A(i,n^\prime)$ and $P_C(i,n^\prime)$ defined in Eqs.~\eqref{eq:ProbA} and~\eqref{eq:ProbC},
\begin{eqnarray}
P_{A}(i,n^\prime)&=&P_{A}(i-1,n^\prime-1)p_{i,1}+P_{B}(i-1,n^\prime-1)p_{i,0}~,\nonumber\\
P_{C}(i,n^\prime)&=&P_{A}(i-1,n^\prime)(1-p_{i,1})\nonumber\\&+&P_{C}(i-1,n^\prime)(1-p_{i,0})~.
\label{eq:RecurProbs1}
\end{eqnarray}
If cell $i$ does not become occupied at time step $t=i$, but cell $i+1$ becomes occupied at some later time-step $t^\prime$ ($i<t^\prime\le N$) then cell $i$ can also become occupied during time-step $t^\prime$, i.e. an avalanche can propagate back down the pore. 
The probability of cell $i$ becoming occupied in this way at time step $t^\prime$ depends upon whether cell $i-1$ is occupied or not.
In fact, if cell $i-1$ is occupied and cell $i+1$ becomes occupied on time step $t^\prime$ then cell $i$ will also become occupied at time step $t^\prime$ 
only if the random field at cell $i$ is in the range $-H<h_i<-H+2J$, which occurs with probability $p_{i,2}-p_{i,1}$. 
On the other hand, if cell $i-1$ is unoccupied then the random field must be in the range $-H-2J<h_i<-H$ in order for cell $i$ to become occupied at time step $t^\prime$ when cell $i+1$ becomes occupied, which occurs with probability $p_{i,1}-p_{i,0}$.
In the case that cell $i-1$ is unoccupied, an avalanche of spin flips can propagate back down the pore from cell $i$ towards cell $1$ during time step $t^\prime$.
This gives the expression for $P_{{B}}(i,n^\prime)$ defined in Eq.~\eqref{eq:ProbB},
\begin{eqnarray}
P_{B}(i,n^\prime)&=&P_{A}(i-1,n^\prime)(1-p_{i,2})\nonumber\\&+&P_{A}(i-1,n^\prime-1)(p_{i,2}-p_{i,1})\nonumber\\
&+&P_{C}(i-1,n^\prime)(1-p_{i,1})\nonumber\\&+&P_{B}(i-1,n^\prime-1)(p_{i,1}-p_{i,0})~.\label{eq:RecurProbs2}
\end{eqnarray}
Eqs.~\eqref{eq:RecurProbs1} and~\eqref{eq:RecurProbs2} lead to the following recursive relations for the generating functions, $A_i(x)$, $B_i(x)$ and $C_i(x)$, 
\begin{eqnarray}
A_i(x)&=&x\left[A_{i-1}(x)p_{i,1}+B_{i-1}(x)p_{i,0}\right]\nonumber\\
B_i(x)&=&x(p_{i,2}-p_{i,1})A_{i-1}(x)+(1-p_{i,2})A_{i-1}(x)\nonumber\\&+&x(p_{i,1}-p_{i,0})B_{i-1}(x)+(1-p_{i,1})C_{i-1}(x)\nonumber\\
C_i(x)&=&(1-p_{i,1})A_{i-1}(x)+(1-p_{i,0})C_{i-1}(x)~,\label{eq:GenRecur}
\end{eqnarray}
valid for $i>1$.

The boundary values of $P_A(1,n^\prime)$ and $P_C(1,n^\prime)$ can be found using the following relations,
\begin{eqnarray}
P_A(1,n^\prime)&=&\delta_{n^\prime,1}p^\prime_{1,0}~,\nonumber\\
P_C(1,n^\prime)&=&\delta_{n^\prime,0}(1-p^\prime_{1,0})~,\label{eq:BaseProbs1}
\end{eqnarray}
where $p^\prime_{1,0}$ is the probability that cell $1$ has a positive field (and thus becomes occupied) at the first time step (when all other cells are unoccupied). The value of $P_B(1,n^\prime)$ is given by
the relation,
\begin{eqnarray}
P_B(1,n^\prime)&=&(1-p_{1,1}^\prime)\delta_{n^\prime,0}+(p_{1,1}^\prime-p_{1,0}^\prime)\delta_{n^\prime,1}~,\label{eq:BaseProbs2}
\end{eqnarray}
where $p_{1,1}^\prime-p_{1,0}^\prime$ is the probability that cell $1$ has a negative local field during time-step $t=1$, but the field becomes positive when cell $2$ becomes occupied, and $1-p_{1,1}^\prime$ is the probability that cell $1$ still has a negative local field after cell $2$ becomes occupied.
Eqs.~\eqref{eq:BaseProbs1} and~\eqref{eq:BaseProbs2} result in the following expression for the boundary generating functions, 
\begin{eqnarray}
A_1(x)&=&xp^\prime_{1,0}~,\nonumber\\
B_1(x)&=&x(p^\prime_{1,1}-p^\prime_{1,0})+1-p^\prime_{1,1}~,\nonumber\\
C_1(x)&=&1-p^\prime_{1,0}~.\label{eq:GenBase}
\end{eqnarray}

The generating function given by Eqs.~\eqref{eq:GenEnd}, \eqref{eq:GenRecur} and~\eqref{eq:GenBase} can be written as a matrix equation, 
\begin{equation}
G(x)=\bm \left[\bm A(x)\right]^{\text{T}}\bm M_{N-1}(x)\bm M_{N-2}(x)\ldots\bm M_{2}(x)\left[\bm B(x)\right]~,\label{eq:GenFuncMatrix}
\end{equation}
where,
\begin{eqnarray}
&&\left[\bm A(x)\right]^{\text{T}}=\left(\begin{matrix}xp^\prime_{N,1}+(1-p^\prime_{N,1}),&xp_{N,0}^\prime,&1-p^\prime_{N,0}\end{matrix}\right)~,\nonumber\\
&&\bm M_i(x)=\nonumber\\ &&\left(
\begin{matrix}
xp_{i,1}&xp_{i,0}&0\\
x(p_{i,2}-p_{i,1})+(1-p_{i,2})&x(p_{i,1}-p_{i,0})&1-p_{i,1}\\
1-p_{i,1}&0&1-p_{i,0}
\end{matrix}\right)~,\nonumber\\
&&\left[\bm B(x)\right]^{\text{T}}=\left(\begin{matrix}xp^\prime_{1,0},&x(p_{1,1}^\prime-p_{1,0}^\prime)+1-p^\prime_{1,1},&1-p^\prime_{1,0}\end{matrix}\right)~.\nonumber\\
\end{eqnarray}
Eq.~\eqref{eq:GenFuncMatrix} is the main analytical result of our analysis allowing exact evaluation of $\partial_xG(1)$ and $\partial_{xx}G(1)$. 
The mean and variance of the volume of fluid in the pore can be found for both adsorption and desorption regimes using Eqs.~\eqref{eq:MeanFromGen} and~\eqref{eq:VarFromGen} along with the derivatives of $G(x)$.
Technically, the derivatives of $G(x)$ can be calculated by numerical iteration, i.e. the derivatives of $\bm M_i(x)\bm M_{i-1}(x)\ldots\bm M_{2}(x)\bm B(x)$ can be found in terms of the derivatives of $\bm M_{i-1}(x)\ldots \bm M_{2}(x)\bm B(x)$.

\section{Results for $T=0$}
\label{Sec:Results_T0}
\begin{figure}
\includegraphics[width=6.0cm]{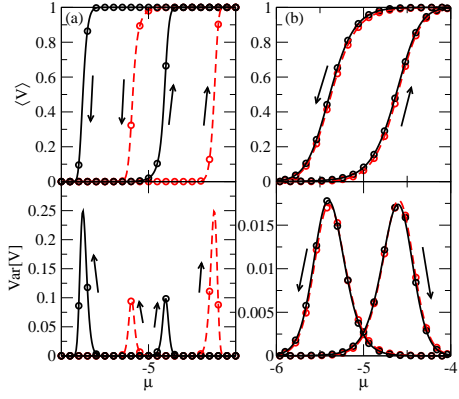}
\caption{Condensation in pores of types I and II (cf. Figs.~\ref{fig:Cell}(a) and (b)). The mean $\langle V\rangle$ (upper panel) and variance $\text{Var}[V]$ (lower panel) of occupied volume of pores of length $N=100$ are plotted vs $\mu$. 
The bold arrows show the direction in which $\mu$ changes for adsorption and desorption curves. 
The solid (dashed) curves correspond to pores of type I (II). 
Different columns refer to different degrees of normal disorder in $w^{\text{mf}}_{ij}$ with the same mean value $\langle w^{\text{mf}}\rangle=1.0$ and the same $w^{\text{ff}}=1.0$: (a) $\rho(w^{\text{mf}})$ has a width $\Delta=0.05$ and (b) a width $\Delta=0.25$.
Symbols refer to the numerical results.
\label{fig:LowFF}}
\end{figure}

\begin{figure}
\includegraphics[width=6.0cm]{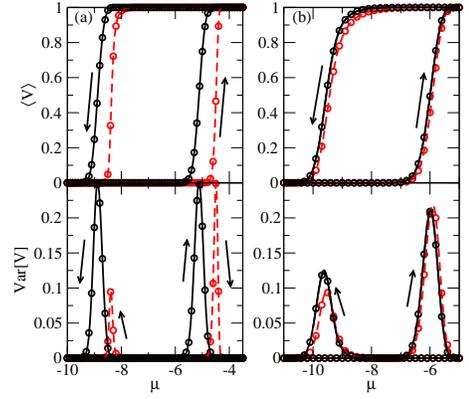}
\caption{Condensation in pores of types I and II. 
As in Fig.~\ref{fig:LowFF}, mean $\langle V\rangle$ (upper panel) and variance $\text{Var}[V]$ (lower panel) of occupied volume of pores of length $N=100$ are plotted vs $\mu$. 
Columns show different values of normal disorder $\Delta$ with constant, $\langle w^{\text{mf}}\rangle=1.0$ and $w^{\text{ff}}=4.0$.
The degree of disorder is $\Delta=0.1$ in column (a) and $\Delta=0.4$ in column (b).
\label{fig:HighFF}}
\end{figure}

The results presented in this section correspond to the numerical solution for $G(x)$ given by Eq.~\eqref{eq:GenFuncMatrix}. 
In order to test the validity of the exact solution, we performed Monte-Carlo simulations of condensation within the framework of the lattice-gas model.
During each simulation, the value of $\mu$ was fixed, and hysteresis plots were obtained by running separate simulations for each value of $\mu$.
The fluid occupation number of the matrix-free cells was changed following single spin-flip zero-temperature Metropolis dynamics~\cite{Sethna1993,Handford2012}.
The mean and variance of $V$ were calculated by averaging over $10^4$ realisations of the disorder in random-fields for a fixed matrix structure and values of the parameters.  
Figs.~\ref{fig:LowFF}, \ref{fig:HighFF}, \ref{fig:ExpDisorder} and~\ref{fig:twoType} show that the analytical calculations (lines) agree with the numerical simulations (symbols). 

The shape of the hysteresis loop and the distribution of $\rho$ for a given value of $\mu$ depends on the pore type and are governed by several parameters.
These parameters are the degree of disorder in matrix-fluid interactions, $\Delta$, the system size, $N$, the relative strengths of matrix-fluid, $\langle w^{\text{mf}}\rangle$, and fluid-fluid interactions, $w^{\text{ff}}$, and the value of the chemical potential, $\mu$. 
The values of the parameters influence the dynamics of the avalanche, i.e. they control the location of the avalanche nucleation points, being either at end points or internal points of the pore, and the number of points at which the avalanches are pinned.
As a result, there are several distinct regimes for sorption in linear pores of all types.

We start out analysis with a detailed description of the different regimes in pores of types I and II.
First, the different possible forms for the sorption curves in the case of a normal distribution in $w^{\text{mf}}_{ij}$ are described.
Second, we analyse the new features which an exponential distribution brings to the sorption curves.
Finally, a description of the sorption curves in pores of type III is presented.

\subsection{Nucleation and Pinning}

When the increasing value of $\mu$ reaches a certain value one of the cells in the system can be filled with liquid, i.e. an avalanche can be nucleated.
This avalanche can propagate either through the whole system or stop at some cell, which becomes a pinning point.
When the avalanche is pinned, a further increase in $\mu$ is required for de-pinning, i.e. for adsorption to continue.
Therefore, there exist two typical values of $\mu$, corresponding to nucleation and de-pinning.
Let us first estimate the value of $\mu$ at which nucleation occurs in pores of both types I and II.

Nucleation of an avalanche occurs at a particular cell $i$ if it is an energetically favourable process, i.e. when the local field given by Eq.~\eqref{eq:LocalField} at that cell is positive.
This happens if the value of chemical potential becomes greater than $\mu_i$, a local nucleation potential, given by,
\begin{equation}
\mu_i(n_i^{\text{m}},n_i^{\text{f}})=-\left(\sum_{j=1}^{n_i^{\text{m}}} w_{ij}^\text{mf}(1-\eta_j)+n_i^\text{f}w^{ff} \right)~,\label{eq:LocNucPot}
\end{equation}
where $n_i^{\text{m}}$ and $n_i^{\text{f}}$ are defined in Secs.~\ref{sec:Mapping} and~\ref{sec:Dynamics}, respectively.
In pores of type I, there are two kinds of cell, end cells, ($i=1$ and $i=N$) and interior cells ($1<i<N$).
For end cells, number of surrounding matrix cells ($n_1^{\text{m}}=n_N^{\text{m}}=5$) is greater than for inner cells ($n_i^{\text{m}}=4$) favouring nucleation at the end cells of type-I pores.
In contrast, all cells in pores of type-II are equivalent for nucleation events because $n_i^{\text{m}}=4$ for $1\le i\le N$.

The values of $\mu_i$ are random as a consequence of the disorder in matrix-fluid interaction strengths $w^{\text{mf}}_{ij}$. 
First, we analyse the case of a normal distribution of $w^{\text{mf}}_{ij}$ given by Eq.~\eqref{eq:gaussDistribution}.
For a quenched configuration of matrix-fluid interaction there will be a cell with a minimal value of $\mu_i$, and the first nucleation event will occur at that cell.
In a pore of type II, this minimal value has a mean, $\left\langle\mu_{\text{min}}^{\text{o}}\right\rangle$, which 
can be estimated using the relation $\text{Prob}[\mu_i\le\left\langle\mu_{\text{min}}^{\text{o}}\right\rangle]\simeq 1/N$, for the mean minimum of $N$ independently and identically distributed random values of $\mu_i$ (see e.g. Ref.~\onlinecite{Sornette2000}).
For a normal distribution of $w^{\text{mf}}$ (see Eq.~\eqref{eq:gaussDistribution}), the estimate is given by the following equation,
\begin{equation}
\left\langle\mu_{\text{min}}^{\text{o}}\right\rangle\simeq -\left[n^{\text{m}}_i\langle w^\text{mf} \rangle + \Delta \sqrt{2n^{\text{m}}_i} \, \text{erfc}^{-1}(2/N) \right]~,\label{eq:OpenNucleation}
\end{equation}
where $n^{\text{m}}_i=4$ for all $i$. 
Here, $\text{erfc}^{-1}(x)$ is the inverse of the complimentary error function, $\text{erfc}(x)=\frac{2}{\sqrt{\pi}} \int_x^\infty e^{-t^2} \text{d}t$. 
As follows from Eq.~\eqref{eq:OpenNucleation}, the value of $\left\langle\mu_{\text{min}}^{\text{o}}\right\rangle$ linearly decreases with increasing degree of disorder (see the dashed lines in Fig.~\ref{fig:Cases}(a) and (b) for the dependence of $\left\langle\mu_{\text{min}}^{\text{o}}\right\rangle$ on $\Delta$) and decreases with pore length $N$ according to $\left\langle\mu_{\text{min}}^{\text{o}}\right\rangle \sim \sqrt{\ln{N}}$. 

For a pore of type I, the values of $\mu_i$ are distributed differently for end and inner cells. 
As a consequence there are two expressions for the mean minimal value of $\mu$ for end and inner cells.
The values of $\mu_1$ and $\mu_N$ are independently and identically distributed 
and the mean of their minimum can be found, in the case of normally distributed values of $w^{\text{mf}}$, according to the exact formula,
\begin{equation}
\left\langle\mu_{\text{min}}^{\text{end}}\right\rangle=-n^{\text{m}}_1\langle w^{\text{mf}}\rangle-\Delta\sqrt{\frac{n^{\text{m}}_1}{\pi}}~,
\end{equation}
with $n^{\text{m}}_1=n^{\text{m}}_N=5$ (see the black lines in Fig.~\ref{fig:Cases}(a) and (b) for $\Delta\alt\Delta^*$).
For inner cells, the expression for the mean value of the minimum, $\left\langle\mu_{\text{min}}^{\text{inner}} \right\rangle$, is given by Eq.~\eqref{eq:OpenNucleation} with $N$ replaced by $N-2$, i.e. for large values of $N$,
\begin{equation}
\left\langle\mu_{\text{min}}^{\text{inner}} \right\rangle\simeq \left\langle\mu_{\text{min}}^{\text{o}} \right\rangle~.
\end{equation}
If $\left\langle\mu_{\text{min}}^{\text{end}} \right\rangle< \left\langle\mu_{\text{min}}^{\text{inner}} \right\rangle$ which is true for $\Delta<\Delta^*=\langle w^{\text{mf}}\rangle\left[\sqrt{8}\,\text{erfc}^{-1}(2/N)-\sqrt{5/\pi}\right]^{-1}$, nucleation starts typically at the end points, and the mean of the minimal value of $\mu_i$ in type-I pores, $\left\langle\mu_{\text{min}}^{\text{c}} \right\rangle$, coincides with $\left\langle\mu_{\text{min}}^{\text{end}} \right\rangle$.
Otherwise, if $\Delta>\Delta^*$, nucleation starts at any cell in the pore with approximately equal probability and $\left\langle\mu_{\text{min}}^{\text{c}} \right\rangle\simeq\left\langle\mu_{\text{min}}^{\text{inner}} \right\rangle<\left\langle\mu_{\text{min}}^{\text{end}} \right\rangle$ (see the coinciding dashed and solid lines Fig.~\ref{fig:Cases}(a) and (b) for $\Delta\agt\Delta^*$).

Once an avalanche is nucleated, it starts propagating. 
This propagation can be stopped by unfavourable variations in the matrix-fluid interaction strength, i.e. it can be pinned at a certain cell.
Pinning will occur at an interior cell $i$ if it is not energetically favourable for cell $i$ to become occupied, even when the avalanche causes one of the cells neighbouring cell $i$ to become occupied by liquid, i.e. pinning occurs at cell $i$ when the random value $\mu_i(4,1)>\mu$.
Typically, pinning will occur when $\mu<\mu_{\text{max}}^{\text{pin}}$ where $\mu_{\text{max}}^{\text{pin}}$ is the mean maximal value of $\mu_i$ for $1<i<N$, which can be found using arguments similar to those for Eq.~\eqref{eq:OpenNucleation} as,
\begin{eqnarray}
&&\mu_{\text{max}}^{\text{pin}}\simeq-\left[n^{\text{m}}_i\langle w^\text{mf} \rangle +w^{\text{ff}}n^{\text{f}}_i- \Delta \sqrt{2n^{\text{m}}_i} \, \text{erfc}^{-1}\left(\frac{2}{N}\right) \right]~,\nonumber\\\label{eq:Pinning}
\end{eqnarray}
with $n^{\text{m}}_i=4$ and $n^{\text{f}}_i=1$. 
The value of $\mu_{\text{max}}^{\text{pin}}$ does not depend on the type of the pore and increases with disorder (see the dot-dashed lines in Fig.~\ref{fig:Cases}(a) and (b)).

\subsection{Regimes for Avalanches}

\begin{figure}
\includegraphics[width=8.5cm]{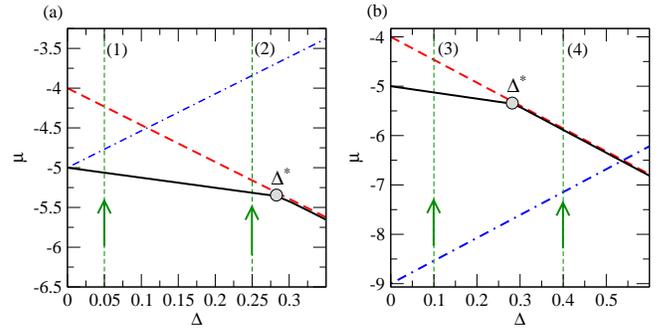}
\caption{Characteristic values of $\mu$ relevant to adsorption for a range of strengths $\Delta$ of normal disorder. 
Shown are the values of $\left\langle\mu^{\text{c}}_{\text{min}}\right\rangle$ (solid line), $\left\langle\mu^{\text{o}}_{\text{min}}\right\rangle=\left\langle\mu^{\text{inner}}_{\text{min}}\right\rangle$ (dashed line) and $\mu^{\text{pin}}_{\text{max}}$ (dot-dashed line). In both panels, $\langle w^{\text{mf}}\rangle=1.0$ and $N=100$. 
The fluid-fluid interaction is $w^{\text{ff}}=1.0$ in panel (a) and $w^{\text{ff}}=4.0$ in panel (b).
The vertical lines with upward arrows indicate values of $\Delta$ corresponding to the four different regimes, (1)-(4), for adsorption described in the text.  
\label{fig:Cases}}
\end{figure}

First, we analyse the different adsorption regimes of adsorption for pores of type I.
The relative values of $\mu_{\text{max}}^{\text{pin}}$, $\left\langle\mu_{\text{min}}^{\text{c}} \right\rangle$ and $\left\langle\mu_{\text{min}}^{\text{inner}} \right\rangle$
(with $\left\langle\mu_{\text{min}}^{\text{c}} \right\rangle \le\left\langle \mu_{\text{min}}^{\text{inner}} \right\rangle$) 
depend on $w^{\text{ff}}$, $\Delta$ and $N$, and four different regimes of adsorption exist.
The first regime is defined by the following sequence of characteristic chemical potentials,
\begin{equation}
\left\langle\mu_{\text{min}}^{\text{c}}\right\rangle<\mu_{\text{max}}^{\text{pin}}<\left\langle\mu_{\text{min}}^{\text{inner}}\right\rangle~.
\label{eq:Regime1}
\end{equation}
An example of adsorption in this regime is shown by the vertical dashed line marked with (1) in Fig.~\ref{fig:Cases}(a), which first crosses the solid line, then the dot-dashed line and finally the dashed line.
This regime can exist for small enough $w^{\text{ff}}$ (e.g. $w^{\text{ff}}=1.0$ in panel (a) of Fig.~\ref{fig:Cases}) such that the crossing point between the dot-dashed line (see Eq.~\eqref{eq:Pinning}) and the dashed line (see Eq.~\eqref{eq:OpenNucleation}) occurs at $\Delta<\Delta^*$, and for a certain range of $\Delta$, e.g. at $\Delta=0.05$ in Fig.~\ref{fig:Cases}(a).
The adsorption in this regime (see solid line in Fig.~\ref{fig:LowFF}(a) upper panel) begins for $\mu\simeq\left\langle\mu_{\text{min}}^{\text{c}}\right\rangle<\left\langle\mu_{\text{min}}^{\text{inner}}\right\rangle$, implying that nucleation typically occurs at the end of the pore.
The variance takes a relatively small value as compared to its maximum possible value of $\text{Var}[V]_{\text{max}}= 0.25$ corresponding to a bi-modal distribution of $V$ with equally probable values $V=0$ and $V=1$\cite{Neri2011}
(see solid line in Fig.~\ref{fig:LowFF}(a) lower panel), meaning that the avalanches nucleated at the ends of the pore progress gradually along the pore between several pinning points with no large sudden jumps in density.
This is in contrast to cases when the varience takes its maximum value, which correspond to a single large avalanche (see regimes (3) and (4) below).

In the regime of large disorder in $w^{\text{mf}}_{ij}$ (regime (2)), pinning dominates over nucleation, and adsorption takes place in the form of many small avalanches nucleated mainly in the inner cells.
An example of adsorption in regime (2) is shown in Fig.~\ref{fig:Cases}(a) by a vertical dashed line, with the following sequence of crossing points, 
\begin{equation}
\left\langle\mu_{\text{min}}^{\text{c}}\right\rangle<\left\langle\mu_{\text{min}}^{\text{inner}}\right\rangle<\mu_{\text{max}}^{\text{pin}}~.
\nonumber
\end{equation}
In this example, adsorption 
can be nucleated first at the closed ends, when $\mu\simeq \left\langle\mu_{\text{min}}^{\text{c}}\right\rangle$.
The avalanche that occurs as a result of such a nucleation is small because of the large number of pinning points which exist for $\mu<\mu_{\text{max}}^{\text{pin}}$.
At a higher $\mu\agt\left\langle\mu^{\text{inner}}\right\rangle$, many more small avalanches are nucleated in the middle of the pore, in contrast to regime (1).
The resulting filling process is gradual, occurring mainly over the range of $\mu$, $\left\langle\mu^{\text{inner}}\right\rangle\alt\mu\alt\mu_{\text{max}}^{\text{pin}}$ (see solid line in upper panel of Fig.~\ref{fig:LowFF}(b)), and characterised by a low variance (see solid line in lower panel).
In fact, if the dot-dashed line is above the others, i.e.
\begin{equation}
\left\langle\mu_{\text{min}}^{\text{inner}}\right\rangle<\mu_{\text{max}}^{\text{pin}}~,\label{eq:Regime2}
\end{equation}
then the dominant effect for adsorption is pinning, meaning that adsorption is in regime (2).
Such a regime thus occurs for large enough $\Delta$ or $N$ for any value of $w^{\text{ff}}$,
i.e. when the value of 
$\Delta$ is to the right of the crossing point between the dashed and dot-dashed lines in Fig.~\ref{fig:Cases}(a) or (b).

The third and fourth regimes (see Fig.~\ref{fig:Cases}(b)), can be achieved by increasing the value of $w^{\text{ff}}$, corresponding to a downward shift of the dot-dashed line so that the intersection point of the dot-dashed and dashed line occurs at $\Delta>\Delta^*$. 
In regimes (3) and (4), pinning is not important, i.e. $\mu_{\text{max}}^{\text{pin}}$ is the smallest relevant value of $\mu$, and adsorption occurs in a single avalanche.
This is demonstrated by the large peak in $\text{Var}[V]$ ($\simeq 0.25$) for both regimes, see right-hand solid peak-shaped curves in the lower panels of Fig.~\ref{fig:HighFF}(a) and (b) corresponding to regimes (3) and (4), respectively.
The boundary between these two regimes occurs at $\Delta=\Delta^*$.
For regime (3), $\Delta<\Delta^*$ meaning that,
\begin{equation}
\mu_{\text{max}}^{\text{pin}}<\left\langle\mu_{\text{min}}^{\text{c}}\right\rangle<\left\langle\mu_{\text{min}}^{\text{inner}}\right\rangle~.
\label{eq:Regime3}
\end{equation}
In this case, adsorption 
occurs in a single avalanche nucleated at one of the end-cells at $\mu\simeq\left\langle\mu_{\text{min}}^{\text{c}}\right\rangle$ (see right-hand solid curve in Fig.~\ref{fig:HighFF}(a) upper panel). 
For regime (4), $\Delta>\Delta^*$ and,
\begin{equation}
\mu_{\text{max}}^{\text{pin}}<\left\langle\mu_{\text{min}}^{\text{c}}\right\rangle\simeq\left\langle\mu_{\text{min}}^{\text{inner}}\right\rangle~,
\label{eq:Regime4}
\end{equation}
so that adsorption occurs in a single avalanche nucleated at one of the inner cells at $\mu\simeq\left\langle\mu_{\text{min}}^{\text{inner}}\right\rangle$ (see right-hand solid curve in Fig.~\ref{fig:HighFF}(b) upper panel).

The length, $N$, of the pore influences the configuration of the boundaries shown in Fig.~\ref{fig:Cases} by changing the slope of the dashed and dot-dashed lines.
The solid line is independent of $N$ for $\Delta<\Delta^*$.
When $N$ increases, the magnitude of both slopes increases proportionally to $\sqrt{\ln N}$.
As such, for very large $N$, only regime (2) can be observed, with nucleation first occurring at very small values of $\mu$ and adsorption taking place as a series of small avalanches until $\mu$ is very large.

\begin{figure}
\includegraphics[width=8.5cm]{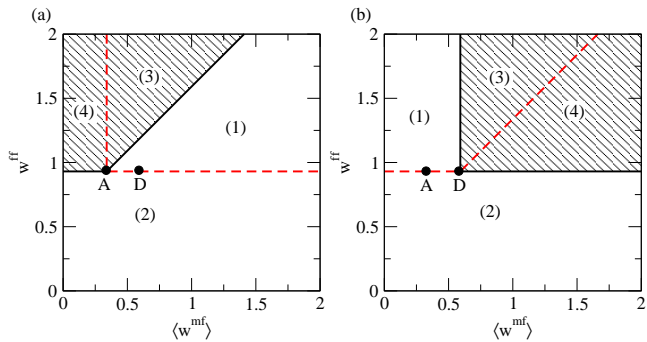}
\caption{Four regimes for (a) adsorption and (b) desorption are shown in the parameter space of $\left\langle w^{\text{mf}}\right\rangle$ and $w^{\text{ff}}$ for pores of type I of length $N=100$ and $\Delta=0.1$.
Points $A$ and $D$, marked on both panels, are given in Eqs.~\eqref{eq:PointA} and~\eqref{eq:PointD}, respectively. Adsorption in pores of type II exhibits two regimes, (2) and (4). Regime (2) corresponds to the region (2) in panel (a) and regime (4) spans over the regions (1), (3) and (4) in (a). Similarly, desorption in pores of type II occurs only in two regimes, (1) and (2). Regime (2) corresponds to the region (2) in panel (b) and regime (1) spans over the regions (1), (3) and (4) in (b).
\label{fig:mfff}}
\end{figure}

The four regimes for adsorption in type-I pores described above can be accessed at constant values of $\Delta$ and $N$ by varying the values of $\left\langle w^{\text{mf}}\right\rangle$ and $w^{\text{ff}}$. 
A diagram showing the boundaries between the regimes (1)-(4) in the parameter space of $\left\langle w^{\text{mf}}\right\rangle$ and $w^{\text{ff}}$ is shown in Fig.~\ref{fig:mfff}(a).
As seen from this diagram, all of these boundaries meet at a point $A\left(\left\langle w^{\text{mf}}\right\rangle,w^{\text{ff}}\right)$ located at,
\begin{equation}
A\left(\Delta\left[\sqrt{8}\,\text{erfc}^{-1}(2/N)-\sqrt{5/\pi}\right],2\sqrt{8}\Delta\text{erfc}^{-1}(2/N)\right)~.\label{eq:PointA}
\end{equation}
For values of $w^{\text{ff}}$ lower than its value at point $A$ the adsorption is in regime (2),
i.e. for such values of $w^{\text{ff}}$ there will be many small avalanches nucleated at inner cells. 
For larger values of $w^{\text{ff}}$ the regime depends on the value of $\left\langle w^{\text{mf}}\right\rangle$.
Regimes (1) and (3), corresponding to avalanches nucleated only at the end-cells of the pore, appear at higher values of both $w^{\text{ff}}$ and $\left\langle w^{\text{mf}}\right\rangle$ than the point A (the region bounded by the dashed line in Fig.~\ref{fig:mfff}(a)).
Regimes (3) and (4), corresponding to avalanches that are not affected by pinning and occur in a single jump, appear at values of $w^{\text{ff}}$ which are higher than both the value at point A, and the value on a line $w^{\text{ff}}=\left\langle w^{\text{mf}}\right\rangle+\Delta\left[\sqrt{8}\,\text{erfc}^{-1}(2/N)+\sqrt{5/\pi}\right]$ which passes through A with gradient $1$ (the shaded region in Fig.~\ref{fig:mfff}(a)). 
To summarise, regime (1): small avalanches nucleated at the end-cells only; regime (2): small avalanches nucleated mainly in the inner cells; regime (3): single large avalanche nucleated at an end-cell; regime (4): single large avalanche nucleated in an inner cell.

For pores of type II, a similar analysis can be performed, by replacing $\left\langle\mu^{\text{c}}_{\text{min}}\right\rangle$ with $\left\langle\mu^{\text{o}}_{\text{min}}\right\rangle$ in Eqs.~\eqref{eq:Regime1}-\eqref{eq:Regime4}.
It can be shown that only two regimes exist, (2) and (4).
The boundary between these two regimes in the parameter space $(\left\langle w^{\text{mf}}\right\rangle, w^{\text{ff}})$ is a line of constant $w^{\text{ff}}$ passing through the same point A as in Fig.~\ref{fig:mfff}(a).
This means that pores of both types are in regime (2) for the same range of parameters.
Conversely, when type-I pores are in regimes (1), (3) or (4), type-II pores are in regime (4).

The shapes of the adsorption curves in different regimes reveal the similarities and differences between the pores of types I and II.
For the same set of parameters, adsorption can be in regime (1) for a type-I pore while it is in regime (4) for a type-II pore. 
In this case, adsorption in the type-I pore is nucleated at the end-cells and progresses slowly between many pinning points along the pore while adsorption in the type-II pore is nucleated in an inner cell and happens in a single avalanche because there are no pinning points.
The difference between the two pore types is due to the additional interactions between the fluid and the end-cap matrix cell of the type-I pores, which are absent in pores of type II and ensure that the value of $\mu$ at which adsorption starts in the type-I pores is lower than in the type-II pores ($\left\langle\mu_{\text{min}}^{\text{c}}\right\rangle\simeq -5.1\alt\left\langle\mu_{\text{min}}^{\text{o}}\right\rangle\simeq -4.2$, compare solid and dashed lines in Fig.~\ref{fig:LowFF}(a) upper panel).
At the low values of $\mu$ when adsorption starts in the pore of type-I, there are many pinning points, which prevent the occurrence of a single avalanche.
This is supported by the fact that
the maximum value of the variance, $\text{Var}[V]$, is smaller than $0.25$ for the type-I pore, but is approximately equal to $0.25$ for a type-II pore (see the lower panel of Fig.~\ref{fig:LowFF}(a)).

Adsorption in regime (2) is governed by disorder for both types of pore and their adsorption curves practically coincide (see Fig.~\ref{fig:LowFF}(b) both panels).
The relatively small values of variance (see Fig.~\ref{fig:LowFF}(b) lower panels) confirm the presence of many pinning points leading to many small avalanches.

When the adsorption is in regime (3) for the pores of type I, it is in regime (4) for the pores of type II.
This means that both pores exhibit a large avalanche, but it is nucleated at the end-cells in a type-I pore and at an inner cell in a type-II pore.
Adsorption in the type-I pore therefore occurs at a lower value of $\mu$ than in a pore of type II (see Fig.~\ref{fig:HighFF}(a) upper panel).
The variance reaches a peak of $0.25$ for both (see Fig.~\ref{fig:HighFF}(a) lower panel), confirming that adsorption occurs in a single avalanche.
At lower values of $\left\langle w^{\text{mf}}\right\rangle$, adsorption is in regime (4) for both types of pore, so that it is nucleated in an inner cell in both cases and occurs in a single avalanche.
This causes the adsorption curves to practically coincide for both pore types (see Fig.~\ref{fig:HighFF}(b)).

For desorption, a similar analysis can be undertaken, and the results of this analysis are presented by the set of lines to the left of the adsorption curves in Figs.~\ref{fig:LowFF} and~\ref{fig:HighFF}.
Four regions, similar to those for adsorption, exist for desorption in pores of type I (see Fig.~\ref{fig:mfff}(b)). 
These are separated by several boundaries, which cross at
the point $D(\left\langle w^{\text{mf}}\right\rangle,w^{\text{ff}})$ located at,
\begin{equation}
{D}\left(\Delta\left[\sqrt{8}\text{erfc}^{-1}(2/N)+\sqrt{5/\pi}\right],2\sqrt{8}\Delta\text{erfc}^{-1}(2/N)\right)~.\label{eq:PointD}
\end{equation}
For values of $w^{\text{ff}}$ lower than that at point $D$ (regime (2)), desorption exhibits many small avalanches nucleated in the inner cells, as in regime (2) for adsorption.
For $w^{\text{ff}}$ higher than that at the point $D$, there are several regimes.
Regimes (3) and (4) correspond to desorption taking place in a single avalanche, which occurs 
when both $w^{\text{ff}}$ and $\left\langle w^{\text{mf}}\right\rangle$ are greater than their values at point $D$ (shaded region in Fig.~\ref{fig:mfff}(b)). 
Regimes (1) and (3) correspond to desorption which is nucleated at the end of the cell, and occur for values of $w^{\text{ff}}$ greater than that at $D$ and greater than a line $w^{\text{ff}}=\left\langle w^{\text{mf}}\right\rangle+\Delta\left[\sqrt{8}\text{erfc}^{-1}(2/N)-\sqrt{5/\pi}\right]$ which passes through point $D$ with gradient $1$ (the region bounded by the dashed line in Fig.~\ref{fig:mfff}(b)). 
Note that the boundaries between the regimes are, in general, different from those for adsorption.
Larger values of $\langle w^{\text{mf}}\rangle$ encourage (discourage) nucleation of adsorption (desorption) at the end-cells, and also encourage (discourage) pinning to occur by making the adsorption (desorption) process start at a lower value of $\mu$. 

For pores of type-II, the desorption can be either in regime (1) or (2) only.
These two regimes are separated in parameter space $(\left\langle w^{\text{mf}}\right\rangle,w^{\text{ff}})$ by the line at constant $w^{\text{ff}}$ passing through points $A$ and $D$.
As such, when the parameters are chosen in such a way that desorption in type-II pores is in regime (2), a type-I pore with the same parameters will exhibit desorption in regime (2) also. 
In this case, the desorption curves and variances coincide for the two pore types (see Fig.~\ref{fig:LowFF}(b)).
For parameters such that desorption in a pore of type II is in regime (1), a type-I pore can be in either types (1), (3) or (4) and the desorption curves, in general, do not coincide.

\subsection{Exponential disorder in matrix-fluid interaction strength}
\begin{figure}
\includegraphics[width=6.0cm]{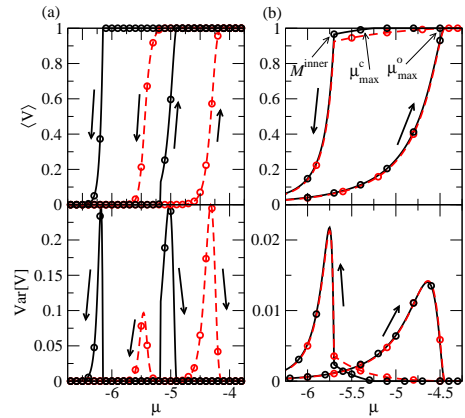}
\caption{Condensation for linear pores of length $N=100$ with exponential disorder of strength $\Delta=0.02$ (panel (a)) and $\Delta=0.2$ (panel (b)). Same line styles are used as in Figs.~\ref{fig:LowFF} and~\ref{fig:HighFF}.
The interaction parameters are $\left\langle w^{\text{mf}}\right\rangle=1.0$ and $w^{\text{ff}}=1.25$.
\label{fig:ExpDisorder}}
\end{figure}

The above analysis has been done for the case of a normal distribution in $w^{\text{mf}}_{ij}$.
The overall picture is qualitatively the same for a correlated exponential distribution of $w^{\text{mf}}_{ij}$ given by Eq.~\eqref{eq:expDistribution}.
However, the presence of a well-defined lower bound for $w^{\text{mf}}_{ij}$ associated with the sharp cut-off in the distribution at $w^{\text{mf}}_{\text{min}}=\left\langle w^{\text{mf}}\right\rangle-\Delta$ leads to two important differences between the two types of disorder. 
Indeed, the sorption curves for correlated exponential disorder display a number of cusp singularities (discontinuities in the derivative of $\langle V\rangle$ with respect to $\mu$) that contrast with the smooth curves for normal disorder in $w^{\text{mf}}_{ij}$ (see Fig.~\ref{fig:ExpDisorder}). 
The second difference is that, for large system size, the limiting behaviour of the sorption is not the same for both types of heterogeneity, i.e. the hysteresis curves for the exponential disorder are asymmetric and of the H2-type~\cite{Sing1985}, in contrast to the parallel-sided H1-type seen for a normal distribution of $w^{\text{mf}}_{ij}$.

\begin{figure}
\includegraphics[width=8.5cm]{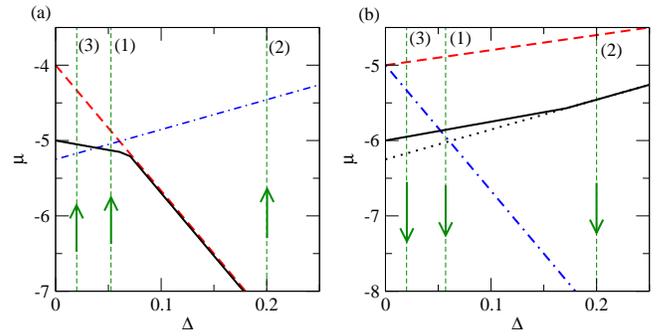}
\caption{Characteristic values of $\mu$ for (a) adsorption and (b) desorption for exponential disorder as functions of the degree of disorder, $\Delta$, for a pore of length $N=100$ with $\left\langle w^{\text{mf}}\right\rangle=1.0$ and $w^{\text{ff}}=1.25$.
In panel (a), solid, dashed and dot-dashed lines indicate $\left\langle\mu^{\text{c}}_{\text{min}}\right\rangle$, $\left\langle\mu^{\text{o}}_{\text{min}}\right\rangle=\left\langle\mu^{\text{inner}}_{\text{min}}\right\rangle$ and $\mu^{\text{pin}}_{\text{max}}$, respectively. 
The vertical lines with upward arrows give examples of adsorption in regimes (1), (2) and (3).
Panel (b) shows $\langle \mu_{\text{max}}^{\text{c}} \rangle$ (solid line), $\langle \mu_{\text{max}}^\text{o} \rangle$ (dashed line), $\mu^{\text{pin}}_{\text{min}}$ (dot-dashed line) and $\langle \mu_\text{max}^\text{inner} \rangle$ (dotted lone).
The dashed curve in panel (b) merges with the dotted and solid curves for large $\Delta$ (off the right-hand edge of the graph). 
Examples of desorption occurring in regimes (1), (2) and (3) are indicated by vertical lines with downward arrows in panel (b).
\label{fig:CasesExp}}
\end{figure}

Similarly to the case of a normal distribution, certain values of $\mu$ can be found using Eq.~\eqref{eq:LocNucPot} at which nucleation and de-pinning typically occur for both adsorption and desorption.
For adsorption,  the values of $\left\langle\mu^{\text{c}}_{\text{min}}\right\rangle$, $\left\langle\mu^{\text{o}}_{\text{min}}\right\rangle$ and $\mu^{\text{pin}}_{\text{max}}$ can be found in a similar way
as for a normal distribution and the resulting dependence of these characteristic values of $\mu$ on $\Delta$ is shown in
Fig.~\ref{fig:CasesExp}(a).

For the case of desorption (see Fig.~\ref{fig:CasesExp}(b)), nucleation will typically occur at the end-cells for $\mu\simeq \left\langle\mu^{\text{end}}_{\text{max}}\right\rangle$,
\begin{equation}
\left\langle\mu^{\text{end}}_{\text{max}}\right\rangle=-n_1^{\text{m}}\left\langle w^{\text{mf}}\right\rangle-w^{\text{ff}}+\frac{n_1^{\text{m}}\Delta}{2}~,\label{eq:EndMaxMean}
\end{equation}
where $n_1^{\text{m}}=n_N^{\text{m}}=5$ for type-I pores (see solid line in Fig.~\ref{fig:CasesExp}(b)) and $n_1^{\text{m}}=n_N^{\text{m}}=4$ for type-II pores (see dashed line in Fig.~\ref{fig:CasesExp}(b)).
For the inner cells, desorption will typically first nucleate at $\mu\simeq\left\langle\mu^{\text{inner}}_{\text{max}}\right\rangle$ (see dotted line in Fig.~\ref{fig:CasesExp}(b)), where,
\begin{equation}
\left\langle\mu^{\text{inner}}_{\text{max}}\right\rangle=-n_i^{\text{m}}\left\langle w^{\text{mf}}\right\rangle-2w^{\text{ff}}+n_i^{\text{m}}\Delta\left(1-\frac{1}{N-2}\right)~,\label{eq:MuEndExpDesorb}
\end{equation}
with the additional factor $2$ before $w^{\text{ff}}$ corresponding to two neighbouring cells being occupied by fluid as opposed to $1$ at the ends and $n_i^{\text{m}}=4$ for $1<i<N$.
For long pores ($N\to\infty$), the gradient of $\left\langle\mu^{\text{inner}}_{\text{max}}\right\rangle$ tends to a limiting value of $n_i^{\text{m}}=4$.
The values of $\left\langle\mu^{\text{end}}_{\text{max}}\right\rangle$ can be, depending on the values of $w^{\text{ff}}$, $N$ and $\Delta$, either greater than or less than $\left\langle\mu^{\text{inner}}_{\text{max}}\right\rangle$ for pores of both types I and II. 
This is in contrast to adsorption, when $\left\langle\mu^{\text{inner}}_{\text{max}}\right\rangle<\left\langle\mu^{\text{end}}_{\text{max}}\right\rangle$ always for pores of type II.

Pinning during desorption can occur only for $\mu$ approximately greater than the mean minimal value of $\mu_i(4,1)$, $\mu\agt\mu^{\text{pin}}_{\text{min}}=\langle\min_{i=2,\ldots,N-1}\mu_i(4,1)\rangle$, i.e. when there is some inner cell which remains occupied when one of its neighbours is unoccupied, thus impeding the propagation of the avalanche.
The value of $\mu^{\text{pin}}_{\text{min}}$ can be calculated and it is equal to,
\begin{equation}
\mu^{\text{pin}}_{\text{min}}=-n_i^{\text{m}}\left\langle w^{\text{mf}}\right\rangle-w^{\text{ff}}-n_i^{\text{m}}\Delta\sum_{n=2}^{N-2}{n}^{-1}~,
\end{equation}
with $n_i^{\text{m}}=4$ (see dot-dashed line in Fig.~\ref{fig:CasesExp}(b)).

There are several regimes for desorption in the case of exponential disorder depending on relative values of the characteristic chemical potentials, $\left\langle\mu^{\text{c}}_{\text{max}}\right\rangle$, $\left\langle\mu^{\text{inner}}_{\text{max}}\right\rangle$, $\left\langle\mu^{\text{inner}}_{\text{max}}\right\rangle$ and $\mu^{\text{pin}}_{\text{min}}$ (see Fig.~\ref{fig:CasesExp}(b)).
If the point where $\left\langle\mu^{\text{c}}_{\text{max}}\right\rangle$ (solid line) merges with $\left\langle\mu^{\text{inner}}_{\text{max}}\right\rangle$ (dotted line) is at higher values of $\Delta$ than the crossing point of $\mu^{\text{pin}}_{\text{min}}$ (dot-dashed line) and $\left\langle\mu^{\text{inner}}_{\text{max}}\right\rangle$, three distinct regimes can be accessed, marked by downward vertical arrows in Fig.~\ref{fig:CasesExp}(b).
In fact, similarly to the case of normally distributed disorder in $w^{\text{mf}}_{ij}$, the parameter space, $\left(\left\langle w^{\text{mf}}\right\rangle,w^{\text{ff}}\right)$ for fixed values of $\Delta$ and $N$, can be split into four regions corresponding to four regimes for either adsorption or desorption.
The boundaries between the different regimes have the same configuration as in Fig.~\ref{fig:mfff} for normal distribution, but are characterised by different locations of the points $A\left(\left\langle w^{\text{mf}}\right\rangle,w^{\text{ff}}\right)$ and $D\left(\left\langle w^{\text{mf}}\right\rangle,w^{\text{ff}}\right)$,
\begin{eqnarray}
&&A\left(\Delta\left[4(1-N^{-1})-\frac{5}{2}\right],4\Delta\sum_{n=1}^{N-3}n^{-1}\right)\nonumber\\
&&D\left(\Delta\left[4(1-N^{-1})+\frac{5}{2}\right],4\Delta\sum_{n=1}^{N-3}n^{-1}\right)
\end{eqnarray}

Below, we analyse two representative sets of parameters.
For the first set of parameters, 
$\left\langle\mu^{\text{o}}_{\text{min}}\right\rangle>\left\langle\mu^{\text{c}}_{\text{min}}\right\rangle>\left\langle\mu^{\text{pin}}_{\text{max}}\right\rangle$, and the type-I and II pores are in regimes (3) and (4), respectively, for adsorption, and in regimes (3) and (1), respectively, for desorption.
The sorption curves for this choice of parameters are presented in Fig.~\ref{fig:ExpDisorder}(a).
It is confirmed that the adsorption in both pore types takes place in a single avalanche.
Typically this avalanche takes place at a smaller value of $\mu$ in type-I pores ($\mu\simeq\left\langle\mu^{\text{c}}_{\text{min}}\right\rangle$) than in type-II pores
($\mu\simeq\left\langle\mu^{\text{o}}_{\text{min}}\right\rangle$).
The desorption also agrees with the prediction, in that it takes place over a range of $\mu$ for the type-II pore around $\mu\simeq\left\langle\mu^{\text{o}}_{\text{max}}\right\rangle$ and in a single large avalanche for the type-I pore at a lower value $\mu\simeq\left\langle\mu^{\text{c}}_{\text{max}}\right\rangle$.

For the second set of parameters, for which $\left\langle\mu^{\text{o}}\right\rangle>\left\langle\mu^{\text{c}}\right\rangle\simeq
\left\langle\mu^{\text{inner}}\right\rangle>\left\langle\mu^{\text{pin}}\right\rangle$, both pore types are in the disorder-controlled regime (2) for adsorption and desorption.
For this set of parameters, the adsorption curves are practically identical (see Fig.~\ref{fig:ExpDisorder}(b)).
The desorption curves, however, differ slightly.
There are several cusps which appear in these sorption curves 
as a consequence of the lower bound for matrix-fluid interaction which are $w^{\text{mf}}_{ij} \geq w^{\text{mf}}_{\text{min}}=\left\langle w^{\text{mf}}\right\rangle-\Delta$. 
The existence of the lower bound to $w^{\text{mf}}_{ij}$ leads to appearance of upper bounds to the 
random variables $\mu^{\text{end}}_{\text{max}}$, $\mu^{\text{inner}}_{\text{max}}$ and $\mu^{\text{pin}}_{\text{max}}$ 
such that,
\begin{eqnarray}
\mu^{\text{end}}_{\text{max}}&\leq&M^{\text{end}}=-n_1^{\text{m}}(\left\langle w^{\text{mf}}\right\rangle-\Delta)-w^{\text{ff}}\label{eq:EndMaxAbs}\\
\mu^{\text{inner}}_{\text{max}}&\leq&M^{\text{inner}}= -n_i^{\text{m}}(\left\langle w^{\text{mf}}\right\rangle-\Delta)-2w^{\text{ff}}\label{eq:MidMaxAbs}\\
\mu^{\text{pin}}_{\text{max}}&\leq&M^{\text{pin}}= -n_i^{\text{m}}(\left\langle w^{\text{mf}}\right\rangle-\Delta)-w^{\text{ff}}~,\label{eq:PinMaxAbs}
\end{eqnarray}
which can be derived from Eq.~\eqref{eq:LocNucPot} by substituting $w^{\text{mf}}_{ij}=w^{\text{mf}}_{\text{min}}=\left\langle w^{\text{mf}}\right\rangle-\Delta$.
Limits are given for desorption at the ends of pores of type I by $\mu^{\text{c}}_{\text{max}}=M^{\text{end}}$ with $n_1^{\text{m}}=5$ and for desorption at the ends of pores of type II by $\mu^{\text{o}}_{\text{max}}=M^{\text{end}}$ with $n_1^{\text{m}}=4$, implying that $\mu^{\text{o}}_{\text{max}}=M^{\text{pin}}$.

For the parameters $\Delta=0.2$ and $w^{\text{ff}}=1.25$ these limiting values are given by, $\mu^{\text{c}}_{\text{max}}=-5.25$, $\mu^{\text{o}}_{\text{max}}=M^{\text{pin}}=-4.45$ and $M^{\text{inner}}=-5.7$.
The location of these limits are marked on the Fig.~\ref{fig:ExpDisorder}(b) by arrows. 
The value of $M^{\text{pin}}$ corresponds to a cusp in the adsorption curves for pores of both types, at which $\langle V\rangle$ rapidly approaches unity and $\text{Var}[V]$ displays a sharp decay to zero.
This cusp occurs because the number of pinning points rapidly approaches zero as the increasing value $\mu\to M^{\text{pin}}$.
For $\mu>M^{\text{pin}}$, there are no pinning points and pore will be fully occupied with probability $1$ if any nucleation event has occurred. 
The mean occupied volume is then close to $1$ in this region.

For desorption, nucleation 
can only occur if $\mu \leq \max\left[{M^{\text{end}},M^{\text{inner}}}\right]$, and as a consequence, 
the occupied volume remains equal to $1$ on decreasing $\mu$ until the above condition is satisfied.
When $\mu$ passes from above to below $M^{\text{end}}$, the probability of nucleation at the ends of the pore begins to increase, giving rise to cusps in the solid and dashed desorption curves at $\mu=\mu^{\text{c}}_{\text{max}}$ and $\mu=\mu^{\text{o}}_{\text{max}}$, respectively (see Fig.~\ref{fig:ExpDisorder}(b)).
At these cusps the value of $\langle V\rangle$ begins to decrease.
The cusps are only significant if the number of pinning points is small enough for the desorption at the end of the cell to cause fluid to desorb in a large part of the pore, meaning that the rate of decrease of $\langle V\rangle$ is large.
A similar effect occurs when $\mu$ passes from above to below $M^{\text{inner}}$, except that the probability of nucleation increases more rapidly with reducing $\mu$, and the resulting cusp in the solid and dashed desorption curves is much sharper, because there are a greater number of inner cells than end cells.
Typically, the cusp at $M^{\text{inner}}$ can only be seen when desorption is in regimes (2) or (4), when nucleation occurs in the inner cells.
Outside of these regimes, desorption fully occurs at higher values of $\mu$ than $M^{\text{inner}}$, and the cusp is insignificant.
For a type-II pore there can be two cusps visible in regime (2), corresponding to nucleation at the end cells and at inner cells.
The desorption curve of a type-I pore can also show two cusps in regime (2) if $M^{\text{inner}}<\mu^{\text{c}}_{\text{max}}$, i.e. if $w^{\text{ff}} \leq \langle w^{\text{mf}} \rangle -\Delta$.
However, if this condition is not satisfied, only one cusp will be visible for a type-I pore, because most desorption will occur as a result of nucleation in the inner part of the pore, i.e. there is no significant cusp at $\mu^{\text{c}}_{\text{max}}$.

The sorption curves for large exponential correlated disorder in $w^{\text{mf}}$ are of the, so-called, H2-type~\cite{Sing1985} (see upper panel of Fig.~\ref{fig:ExpDisorder}(b)).
Such a shape of sorption curves contrasts with that for Gaussian disorder which are of the parallel-sided H1-type (see upper panels of Figs.~\ref{fig:LowFF} and~\ref{fig:HighFF}).
It should be noted that the asymmetry in the distribution of $h_i=\sum_{j/i}w^{\text{mf}}_{ij}$ for the case of exponential disorder is a consequence of a correlated asymmetric distribution of $w^{\text{mf}}_{ij}$.
Indeed, if the values of $w^{\text{mf}}_{ij}$ are uncorrelated for a given value of $i$ then 
the central limit theorem ensures that their sum, $\sum_{j/i}w^{\text{mf}}_{ij}$, is approximately distributed according to a normal distribution, which is symmetric.
Therefore correlations in $w^{\text{mf}}_{ij}$ play a significant role in achieving a skewed distribution of $h_i$ and H2-type hysteresis, the effect being maximal when they are fully correlated, i.e. when all $w^{\text{mf}}_{ij}$ are equal to each other for a given $i$.
This implies that H2-type hysteresis in heterogeneous pores might arise due to variations in pore diameter (represented by correlated disorder in $w^{\text{mf}}_{ij}$) rather than due to individual defects, in agreement with previous numerical studies~\cite{Naumov2008}. 
On the other hand, symmetric (normal) disorder in local fields (i.e. uncorrelated disorder in $w^{\text{mf}}_{ij}$) can only cause a parallel sided H1-type hysteresis (Fig.~\ref{fig:LowFF} and~\ref{fig:HighFF}), and may represent the effect of uncorrelated structural defects in the pore surface on small length scales. 

\begin{figure}
\includegraphics[width=8.5cm]{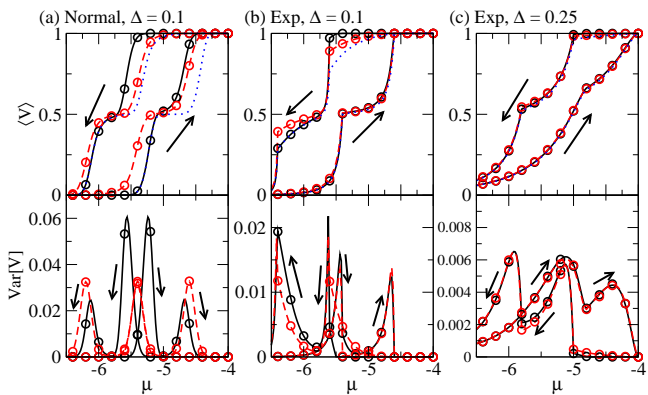}
\caption{Condensation in pores of type III (cf. Fig.~\ref{fig:Cell}(c)). $\langle V\rangle$ (upper panel) and $\text{Var}[V]$ (lower panel) plotted vs $\mu$ for pores of length $N=100$.
The curves of different styles refer to different geometries, i.e. ink-bottle (solid lines), with $N_1=50$, $w_1=1.2$, $w_2=1.0$ and funnel (dashed lines), with $N_1=50$, $w_1=1.0$, $w_2=1.2$.
The light dotted lines in the upper panels of (a)-(c) correspond to the mean occupied volume of two separate open-ended pores (shape (b) in Fig.~\ref{fig:Cell}) of length $N=50$ one of which having $\langle w^{\text{mf}}_{ij}\rangle=1.0$ for all cells and the other having $\langle w^{\text{mf}}_{ij}\rangle=1.2$ for all cells.
In (c), the solid, dashed and dotted lines coincide on the scale of the graph. 
In all cases, $w^{\text{ff}}=1.0$. 
Each column represents a different degree or form of disorder: (a) normally distributed, $\Delta=0.1$
(b) correlated exponentially distributed, $\Delta=0.1$ and (c) correlated exponentially distributed, $\Delta=0.25$.
Arrows show the direction of change of $\mu$ for adsorption and desorption and symbols refer to numerical data. \label{fig:twoType}}
\end{figure}

\subsection{Pores of type III}
In this section, we present results on the effects of interaction between the two parts of a pore, each of length $N_1=N_2=50$, characterised by different mean matrix-fluid interaction strength $\langle w_1^{\text{mf}}\rangle$ and $\langle w_2^{\text{mf}}\rangle$ (as shown in Fig.~\ref{fig:Cell}(c)).
The difference in the interaction strength of the fluid with the matrix can represent variable pore diameter for a pore of either a funnel or an ink-bottle shape.
A major feature of sorption in such pores is that, if the difference between $\langle w_1^{\text{mf}}\rangle$ and $\langle w_2^{\text{mf}}\rangle$ is large enough, it occurs in the narrow part of the pore at a lower value of $\mu$ than it does in the wider part.
This gives rise to two steps in the sorption curves (see upper panels of Fig.~\ref{fig:twoType}), in agreement with experimental observations~\cite{Wallacher2004,Grosman2011} for both pore shapes. 
In order to understand the form of the sorption curves in such pores, it is helpful to compare it with 
the mean of the sorption curves for two independent pores of type II of length $N_1=N_2=N=50$, one with $w^{\text{mf}}=\langle w_1^{\text{mf}}\rangle$ and the other with $w^{\text{mf}}=\langle w_2^{\text{mf}} \rangle$ (see dotted lines in upper panel of Fig.~\ref{fig:twoType}).
The sorption curves (solid and dashed lines) differ from the dotted lines representing the behaviour of two independent type-II pores.
These differences are due to the interaction between fluid in the two parts of the pore and of fluid with the closed end-cap of the type-III pore.

We analyse these differences, first, for a pore with normal disorder in $w^{\text{mf}}_{ij}$ (see dashed lines in Fig.~\ref{fig:twoType}(a)) and a funnel shape, characterised by $\langle w_1^{\text{mf}}\rangle<\langle w_2^{\text{mf}}\rangle$, i.e. the matrix-fluid interaction is weaker in the part of the pore with the larger diameter. 
Adsorption in each part of the pore is enhanced in comparison with a type-II pore (the dashed line corresponding to adsorption is shifted to the left by around $0.25$ with respect to the dotted line for the two steps).
The shift of the lower step (at $\mu\simeq -5.5$) is due to the increased matrix-fluid interaction between the narrow section and the closed end and the shift of the upper step (at $\mu\simeq -4.75$) is due to the fluid-fluid interaction between the wide section and the narrow section.
Conversely, desorption in the funnel pore occurs later than in two independent open-ended pores (the dashed line corresponding to desorption is shifted to the left by around $0.1$ with respect to the dotted line).
This is because the funnel pore has only one open end at which desorption can nucleate and thus it occurs later than in a pore open at both ends.

Second, we compare the sorption curves for an ink-bottle pore with the sorption curves of two independent pores (cf. solid and dotted lines in Fig.~\ref{fig:twoType}(a), respectively).
Both the adsorption and desorption curves for the narrow part of the ink-bottle pore closely match those for an open ended pore of the same diameter (the dotted and solid lines practically coincide in the lower step until the shoulder develops).
Qualitatively, for this part of the adsorption and desorption curves, the wider part of the pore is not occupied by the fluid, meaning that the narrow part effectively has two open ends. 
The upper step in the sorption curves, after the shoulder, represent adsorption and desorption in the wide part of the ink-bottle pore and differ from an open ended pore (cf. the upper steps in the solid and dotted lines in Fig.~\ref{fig:twoType}(a)).
This is because sorption in the wide part occurs when the narrow part is fully occupied with fluid, meaning that 
the wide part of the ink-bottle pore behaves like a pore closed at both ends.
As such, adsorption occurs for smaller values of $\mu$ (the upper step in the solid adsorption curve is shifted to the left by around $0.25$ compared to the dotted line), since adsorption is nucleated either by the closed end or by the fluid which has already condensed in the narrow part.
Similarly, desorption occurs at a lower value of $\mu$, since it is not nucleated at either end (the upper step in the solid desorption curve is also shifted by around $0.25$ to the left compared to the dotted line).
It can be noted that the behaviour on desorption observed here for ink-bottle pore and funnel pore topologies is similar to that experimentally observed in Ref.~\onlinecite{Grosman2011}.
The adsorption curves observed here differ from the experimental results of Ref.~\onlinecite{Grosman2011}, in that condensation in the wider part of an ink-bottle pore is observed at higher values of $\mu$ than in the wider part of a funnel pore.
This might be related to the fact that, in our model, condensation is allowed to occur in the inner part of the ink-bottle pore regardless of the fact that gas has no way to flow into the wider part of the pore (the pore is blocked by fluid in the narrow part)~\cite{Casanova2012}. 

Results for exponential disorder in $w^{\text{mf}}_{ij}$ are shown in Fig~\ref{fig:twoType}(b). 
The dotted line corresponds to a pore with two open ends and the dashed line corresponds to a funnel pore, which effectively has a single open end. 
The solid line shows the isotherms for the ink-bottle pore where, effectively, the wide part has two closed ends and the narrow part has two open ends.
As such, the main differences between the different sorption curves are similar to the differences observed between open- and closed-ended pores in Fig.~\ref{fig:ExpDisorder}(b), i.e. the adsorption curves are broadly similar for both funnel and ink-bottle pores.
The main differences between the isotherms for different pore geometries is observed in two intervals of $\mu$: $-5.6<\mu<-4.6$ and $-6.5\alt \mu \alt -6$. 
In the first interval, desorption can be nucleated at the open end of the wide part of the funnel pore, but nucleation cannot occur in the wide part of the ink-bottle pore because this part effectively has both ends closed.
This is why the dashed line is below the solid line in this region.
In the second interval, the wide parts of both the funnel and ink-bottle pores are empty.
This means that the narrow part of the funnel-shaped pore effectively has one open end while the narrow part of the ink-bottle pore effectively has two open ends.
Desorption is therefore more likely to be nucleated in the narrow part of the ink-bottle shaped pore than in the narrow part of the funnel pore.
Correspondingly, the dashed line is above the solid line in this region.

In the disorder-controlled regime (see Fig.~\ref{fig:twoType}(c)), 
the sorption curves for both funnel and ink-bottle pores are identical to those found by adding together the sorption curves for two independent open-ended pores of different diameters obtained by cutting the funnel and ink-bottle pores in half, i.e. the effects of the ends of the pores are small for disorder on this scale. 
This behaviour appears to correspond to the experimental observations in Ref.~\onlinecite{Wallacher2004}.

\section{Results for $T>0$}
\label{Sec:Results_Tgt0}
\begin{figure}
\includegraphics[width=8cm]{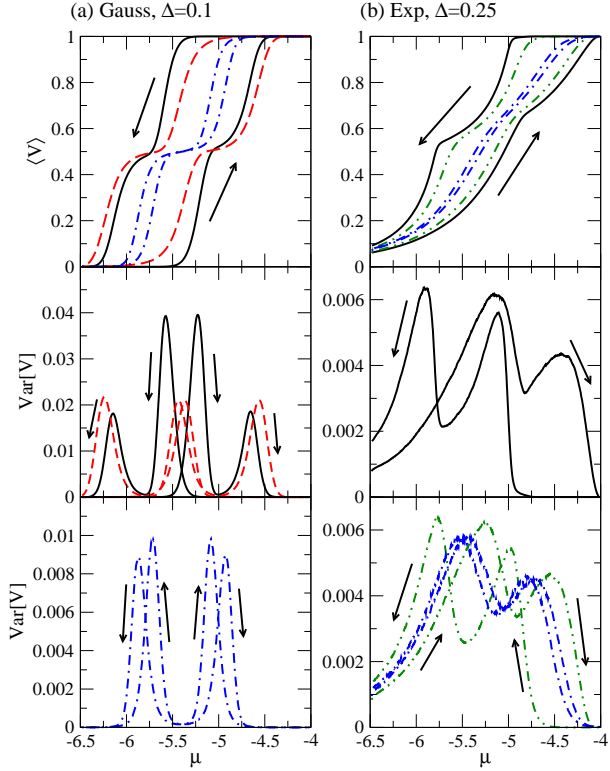}
\caption{Adsorption/desorption in pores of type III at non-zero temperature. Upper panel: Volume of adsorbed fluid {\it vs} $\mu$ for Metropolis dynamics simulations. 
In panel (a), the system consists of $N=100$ cells with ink-bottle shape (solid and dot-dashed lines) and funnel shape (dashed lines) with $\langle w_1^{\text{mf}}\rangle=w^{\text{ff}}=1.0$, $\langle w_2^{\text{mf}}\rangle=1.2$ and Gaussian disorder of strength $\Delta=0.1$. 
Two values of temperature are considered: $\beta=50.0$ (solid and dashed lines) and $\beta=5$ (dot-dashed lines). 
In panel (b), the system has an ink-bottle shape with $\langle w_1^{\text{mf}}\rangle=w^{\text{ff}}=1.0$, $\langle w_2^{\text{mf}}\rangle=1.2$ and correlated exponential distribution of $\rho(w^{\text{mf}})$ with $\Delta=0.25$ and three values of temperature: $\beta=50$ (solid curves), $\beta=15$ (double-dot dashed curves) and $\beta=5$ (dot-dashed curves). 
Middle and lower panels: The variance $\text{Var}[V]$ {\it vs} $\mu$ for the same systems as in the upper panels (with the same line styles).
For clarity, the curves for $\beta=50$ have been presented on the middle panel while curves corresponding to $\beta=15$ and $\beta=5$ are on the lower panel.
In all of the simulations, the rate of change of $\mu$ was $r=0.004\text{MCSS}^{-1}$.
\label{fig:Metropolis}}
\end{figure}

The effect of finite temperature on sorption processes is analysed here for both a funnel and ink-bottle pore. 
In particular, the conditions corresponding to Figs.~\ref{fig:twoType}(a) and~\ref{fig:twoType}(c) were repeated for several values of temperature and with a constant rate, $r$, of change of $\mu$. 
The results are shown in Fig.~\ref{fig:Metropolis}.

For finite sufficiently low temperature, $\beta=50$, and Gaussian disorder of strength $\Delta=0.1$, it can be seen by comparing Figs.~\ref{fig:twoType}(a) and~\ref{fig:Metropolis}(a) that the mean volume of condensed fluid closely matches that at zero-temperature for both ink-bottle and funnel pores.
The main difference from the zero-temperature and small $r$ limit is that the meniscus propagates slowly along the pore rather than the whole pore becoming occupied simultaneously, reflected in a reduction in the maximum value of $\text{Var}[V]$. For instance, the maximum value of the variance for an ink-bottle shaped pore is $\text{Var}[V]\simeq 0.04$ (see peak of solid line in Fig.~\ref{fig:Metropolis}(a) middle panel) in comparison with $\text{Var}[V]\simeq 0.06$ at zero-temperature (see solid line in Fig.~\ref{fig:twoType}(a), lower panel).
For larger temperatures, $\beta=5$, hysteresis loops become narrower (cf. dot-dashed curve with solid curve in Fig.~\ref{fig:Metropolis}(a)).

For exponential disorder of strength $\Delta=0.25$, the main effect of the finite temperature is in the smoothing of the cusps in the hysteresis loops (see Fig.~\ref{fig:Metropolis}(b) upper panel). 
As the temperature is increased, first to $\beta=15$ (double-dot dashed curves) and then to $\beta=5$ (dot-dashed curve), the area of the hysteresis loop reduces gradually.
This observation with increasing temperature agrees with the behaviour observed for 1D pores\cite{Grosman2011} and is similar to the decreasing width of sorption hysteresis loops in 3D porous media~\cite{Gelb1999}. 
Although the behaviour in 1D and 3D systems is similar, the proposed model and its mapping to the RFIM suggests that the origin of hysteresis is not identical in both cases. 
Indeed, the disorder-temperature phase diagram of a 3D lattice gas (or the 3D RFIM) consists of a phase at low temperature and disorder with ferromagnetic order and a paramagnetic phase at high temperature and disorder~\cite{Cardy1996}.  
In the ferromagnetic phase, the free energy consists of two global minima such that gas and liquid (or states with positive and negative magnetisation in the RFIM) could in principle coexist in the thermodynamic limit\cite{Woo2003}. 
Hysteresis is associated in this case with both the global minima of the free energy and the existence of many local minima where the system can remain trapped for very long times. 
The mapping of the proposed model to the 1D-RFIM indicates that ferromagnetic order does not exist as a stable macroscopic phase at any finite temperature and/or disorder~\cite{Cardy1996}. 
Instead, the free energy exhibits a single global minimum. 
Therefore, hysteresis in linear pores at non-zero disorder and temperature is expected to be only associated with the rugged character of the energy landscape which consists of many local minima.

\section{Conclusions}
\label{Sec:Conclusions}
To conclude, a heterogeneous lattice-gas model has been proposed to describe fluid condensation in 1D pores of different shapes and rough surfaces.
Heterogeneity, missed in classical theories, is the key and sufficient feature of the model which allows it to reproduce the main experimental findings.
We demonstrate that a simple 
coarse-grained representation of pores consisting of 1D chains of cells is a minimal model
sufficient to account for the effects of heterogeneity.
Within a single cell, liquid interacts with the elements of the surface.
These interactions can be either identical or variable (random) within the cell, and also can vary for different cells.
Accounting for such different types of heterogeneity in the model, results in the different shapes of hysteresis loop found experimentally~\cite{Sing1985}, including both the H2-type~\cite{Wallacher2004} and the H1-type~\cite{Casanova2008} hysteresis loops.

In addition, the model is able to reproduce the shape of the sorption curves for some more complex experimentally studied systems, including ink-bottle and funnel shaped pores. 
The physical phenomena inherent for this model include nucleation of adsorption and desorption (cavitation) and propagation of a meniscus through the pore, which are known~\cite{Grosman2011} to be two main effects observed in such systems.
In this respect, our model could be easily extended to account for the fluid blocking effect that prevents flow of gas to the wider part of the ink-bottle structure~\cite{Grosman2011,Casanova2012}.

Besides reproducing experimental observations, our model also suggests interesting predictions that motivate new experiments. 
For instance, we have demonstrated that the sorption mechanisms (i.e. whether it starts at the ends of pores or in the interior part) might depend on the length and diameter of the pores and the degree of heterogeneity. 
However, for large heterogenity, or long pores, the sorption tends to a limiting, and apparently universal, disorder-controlled regime.

In physical systems exhibiting a more complex topology, e.g. a 3D maze-like network of 1D channels, an analytical solution may be nontrivial, however numerical simulations within our model can be performed straightforwardly for porous media of arbitrary topology\cite{Monson_MicroMesoMat2012}.

\section{Acknowledgements}

TPH acknowledges the support of EPSRC.

\end{document}